# Near-forward Raman scattering by bulk and surface phonon-polaritons in the model percolation-type ZnBeSe alloy

R. Hajj Hussein,[1] O. Pagès,[1,*] F. Firszt,[2] W. Paszkowicz,[3] and A. Maillard[4]

[1] *LCP-A2MC, Institut Jean Barriol, Université de Lorraine, France*
[2] *Institute of Physics, N. Copernicus University, 87-100 Toruń, Poland*
[3] *Institute of Physics, Polish Academy of Sciences, 02-668 Warsaw, Poland*
[4] *LMOPS, Université de Lorraine – Sup´elec, 2, rue Edouard Belin, 57070 Metz, France*

We study the bulk and surface phonon-polaritons of the $Zn_{0.67}Be_{0.33}Se$ zincblende alloy by near-forward Raman scattering. The short (Be-Se) bond exhibits a distinct percolation doublet in the conventional backscattering Raman spectra, corresponding to a three-mode behavior in total [1×(Zn-Se),2×(Be-Se)] for $Zn_{0.67}Be_{0.33}Se$. This offers an opportunity to achieve a refined understanding of the phonon-polariton modes of a zincblende alloy beyond the current two-mode approximation, corresponding to a [1×(Zn-Se),1×(Be-Se)] description in the present case. The discussion is supported by contour modeling of the Raman signals of the multi-mode bulk and surface phonon-polaritons within the formalism of the linear dielectric response.

---

[*] Corresponding author: olivier.pages@univ-lorraine.fr

In such a polar material as a zincblende AB crystal, the long-wavelength ($q$~0,$\Gamma$) optic vibrations, corresponding to out-of-phase displacements of the quasi rigid cationic-A and anionic-B intercalated *fcc* sublattices, create a macroscopic electric field.[1] The transverse optic ($TO$) mode, ideally described in terms of an effective A-B stretching perpendicular to the direction of propagation, is particularly interesting in that it gives rise to a transverse electric field, thus identical in nature to that carried by a pure electromagnetic wave, i.e. a photon. One may thus hope that, at least for certain $q$ values, the resulting coupled electromagnetic-mechanical $TO$ mode, known as a phonon-polariton (PP), acquires a pronounced electromagnetic character, and as such propagates at 'lightlike' speeds. This has opened the way for ultrafast (photon-like) signal processing at THz (phonon-like) frequencies.[2] Such coupling is likely to occur close to $\Gamma$, due to the quasi vertical dispersion of a photon at the scale of the phonon Brillouin zone (BZ). The PP's propagating in the bulk or at the surface of various polar compounds have been studied, both experimentally and theoretically, with GaP as a prototype system.[3-9] The main features are summarized at a later stage.

In this work we are interested in the PP's of an alloy. We are aware of only one related experimental work so far, i.e. a detailed study of the surface PP of the wurtzite $Al_xGa_{1-x}N$ alloy by attenuated total reflectance.[10] Bao and Liang tackled the problem on a theoretical basis recently, by calculating the dispersion curves ($\omega$ vs. $q$) of the bulk and surface PP's of various $A_{1-x}B_xC$ zincblende alloys.[11] In doing so they presumed that the purely-mechanical $TO$ phonons behind the PP's of such alloys, as currently detected in a conventional backscattering Raman experiment (see below), obey the modified-random-isodisplacement (MREI) model.[12] In this model the like bonds of a given species are taken as insensitive to their local (B,C)-environment, thus contributing to the same unique Raman line at a given $x$ value. This results in a [1×(A-C),1×(B-C)] two-mode behavior in total for the alloy, with several variants.[12] When $x$ changes, the A-B and A-C modes shift regularly between their respective parent and impurity limits, their intensities scaling as the related $x$ and $(1-x)$ bond fractions.

We have shown in the past decade that the MREI model underestimates by far the natural richness of the backscattering Raman spectra of a zincblende alloy. A better description seems to be achieved within the novel percolation model, which apparently applies universally among such alloys.[13] In brief, this model distinguishes between the two possible AC- or BC-like first-neighbor environments of a bond in a phenomenological description of the crystal at one dimension along the linear chain approximation.[14] This leads to *two* distinct $TO$ phonons per bond at a given $x$ value, hence an overall [2×(A-C),2×(B-C)] four-mode behavior in total for the alloy. In a $\omega$ vs. $x$ plot this comes to divide each MREI branch into two quasi parallel sub-branches. An important parameter is the splitting $\Delta$ between two like sub-branches, characteristic of a given bond in a given alloy. Regarding intensities, the overall MREI trend between the A-B and A-C signals is preserved. However, within each doublet the dominant sub-mode at one end of the composition domain turns minor at the other end, the two sub-modes having similar intensities at $x$~0.5.

In this work we explore, both experimentally and theoretically, the PP's of the model $Zn_{1-x}Be_xSe$ alloy, for which the percolation scheme was originally developed. We use a large (~1 cm³, cylinder) (111)-oriented $Zn_{0.67}Be_{0.33}Se$ single crystal grown from the melt, in which the Zn↔Be substitution was shown to be random (see below), and which exhibits the best-resolved percolation doublet of the $Zn_{1-x}Be_xSe$ series.[15] This relates to Be-Se ($\Delta_{Be-Se}$~50 cm⁻¹). In contrast, a single MREI-like mode is observed for Zn-Se ($\Delta_{Zn-Se}$~0 cm⁻¹). This gives a three-mode ($TO_{Zn-Se}$, $TO_{Be-Se}^{Be}$, $TO_{Be-Se}^{Zn}$) behavior in total for Zn$_{0.67}$Be$_{0.33}$Se, as apparent in its conventional backscattering Raman spectra. In this notation, the subscript refers to a given bond vibration, and the superscript to the cation dominating in the bond's local environment. The three-mode Zn$_{0.67}$Be$_{0.33}$Se alloy offers a chance to achieve a refined understanding of the PP's of a zincblende alloy beyond the current two-mode MREI approximation used by Bao and Liang.[10,11] As Zn$_{0.67}$Be$_{0.33}$Se is transparent in the visible range, we opt for a Raman experiment in a near-forward scattering geometry to access its PP modes.[3] The discussion is supported by a contour modelling of the Zn$_{0.67}$Be$_{0.33}$Se PP Raman lineshapes along the formalism of the linear dielectric response developed by Hon and Faust.[16]

For reference purpose we recall briefly the main features of a PP in the case of a zincblende AB compound, such as GaP.[3] When searching for a plane wave solution, thus propagating in the bulk of the crystal, and accordingly referred to as a bulk PP hereafter, Maxwell's equations dictate a $q$-dependent electric field within some few percent of the BZ size,[1] identifying with the actual PP regime. Beyond this range, the $TO$ mode enters its so-called $q_{inf.}$-regime and asymptotically reduces to a purely mechanical $TO$ phonon. Additional photon-like asymptotes $\omega = \frac{qc}{\sqrt{\varepsilon_r}}$, where $c$ is the speed of light in vacuum and $\varepsilon_r$ is the relative dielectric function of the crystal at frequency $\omega$, are defined by the static $\varepsilon_{stat.}$ ($\omega \ll \omega_{TO}$) and high-frequency $\varepsilon_{inf.}$ ($\omega \gg \omega_{TO}$) relative dielectric constants of the crystal far from the $TO$ resonance. Strong PP coupling occurs when the quasi-vertical photon-like asymptotes ($\varepsilon_r$ is small) and the horizontal $TO$-like ones ($\varepsilon_r$ diverges), taken independently, cross each other. The coupling manifests itself via a strong repulsion of two (curved) PP branches, resulting in an effective anticrossing. The lower branch has a photon character at small frequency ($\omega \ll \omega_{TO}$) and a $TO$ character in the $q_{inf.}$-regime ($\omega \sim \omega_{TO}$), while the upper mode is photon-like at high-frequency ($\omega \gg \omega_{TO}$) and assimilates with the *longitudinal* optical ($LO$) phonon while approaching Γ ($\omega \sim \omega_{LO}$).[3,17] Note that the $LO$ phonon carries a stable, i.e. $q$-independent, electric field.[1] This acts as an additional restoring force,[18] so that the $LO$ phonon vibrates at a higher frequency than the purely-mechanical $TO$ one.[19] From the above discussion the $TO - LO$ band is forbidden for the bulk PP. In fact, $\varepsilon_r$ is negative therein, meaning that only a surface PP is allowed (see below).[5,9,11]

An interesting question is how such picture for the single optic phonon of an AB compound does transpose to our three-phonon Zn$_{0.67}$Be$_{0.33}$Se alloy? A convenient technique to cover such issue is Raman scattering.[3] Depending on the scattering geometry, different $q$ values can be addressed. The scattering geometry is all contained in the wavevector conservation rule $\vec{k}_i - \vec{k}_s = \vec{q}$, in which $\vec{k}_i$ and $\vec{k}_s$ refer to the incident laser (visible) and to the scattered light, respectively, both taken inside the crystal, defining an angle $\theta$ therein. In the usual backward geometry ($\theta$=180°), $\vec{k}_i$ and $\vec{k}_s$ are opposite, so that $q$ is maximum, falling, in fact, deep into the $q_{inf.}$-regime.[3] In the less common forward geometry ($\theta$=0°), $\vec{k}_i$ and $\vec{k}_s$ have the same direction (see a drawing in **Fig. 1**), so that $q$ is minimum, and likely to fall within the strong PP coupling regime.[3] Intermediate $q$ values are accessed by varying $\theta$.

In **Fig. 1** we compare the unpolarized Raman spectra of our Zn$_{0.67}$Be$_{0.33}$Se ingot taken in backward ($\theta$=180°, thick curve) and forward ($\theta$=0°, thin curve) scattering along its [111]-growth axis using the 488.0 nm excitation of a cw Ar+ laser with a laser output of 150 mW. Theoretically forbidden transverse ($DA - TA$) and longitudinal ($DA - LA$) acoustic one-phonon density of states activated by the alloy disorder,[15] are used for intensity normalization. In forward scattering, the $LO_{Zn-Se}$ (~250 cm⁻¹) mode and the dominant $LO^+_{Be-Se}$ (~500 cm⁻¹) component of the Be-Se $LO$ percolation doublet[15] are slightly enhanced due to substantial Raman backscattering at the back surface of the sample (laser side) after reflection of the laser beam inside the crystal at its top surface (detector side).[3] The minor overdamped $LO^-_{Be-Se}$ component (~430 cm⁻¹, see below), as for it, remains screened by the $TO$ mode.[15] The latter mode is most affected, as schematically indicated by arrows at the bottom of **Fig. 1**. When turning from backward to forward scattering the Zn-Se (~215 cm⁻¹) $TO$ mode weakens and softens, and also the lower Be-Se $TO$ mode (~420 cm⁻¹), while the upper Be-Se $TO$ mode (~470 cm⁻¹) just disappears.

For a deeper insight into such presumed PP effects, we record near-forward Raman spectra with the 488.0 nm excitation by varying $\theta$. In doing so, we cross the polarizations of the incident ($p$-polarized, $\vec{e}_i$) and scattered ($\vec{e}_s$) beams and rotate the sample until these coincide with the $[1\bar{1}0]$ and $[11\bar{2}]$ crystal axis, corresponding to forbidden $LO$ scattering. The as-obtained $\theta$-dependence of the Be-Se doublet, now fully clarified, is shown in **Fig. 2a**. Similar Be-Se spectra are obtained with the 514.5 nm Ar+ laser line (not shown). The Zn-Se signal (not shown) interferes with the parasitical $DA - LA$ band, as evidenced by a characteristic Fano antiresonance (marked by a star in **Fig. 1**), thus taken as non reliable.[15] The backscattering-like Be-Se Raman signal is preserved down to $\theta$~3.2°. With further reduction of $\theta$, the two Be-Se modes weaken. However, the upper mode weakens faster, so that quasi 1:1 intensity matching is achieved at $\theta$~1°, the original intensity ratio at $\theta$=180° being 1:2. Also, the two modes soften. Remarkably the upper Be-Se mode begins to harden when $\theta$ approaches 0°, eventually stabilizing into the so-called X mode at $\theta$~0° located deep into the $TO - LO$ band, where $\varepsilon_r$ is negative (see **Fig. 3**). Such in-band incursion is forbidden for a bulk PP, suggesting that X is a surface PP.

For a direct insight, we fix $\theta$~0° and defocus progressively the lens collecting the scattered light from its original position, coinciding with the back surface of the sample, where the 488.0 nm laser beam remains focused. The idea is to probe scattering volumes deeper and deeper into the bulk ingot, where the surface PP cannot exist, by definition, and should be replaced by the bulk PP. As expected, the upper Be-Se mode, originally identifying with X, backshifts outside the $TO - LO$ band when defocusing, until full recovery of the bulk PP, as shown in **Fig. 2b**. Note that the $LO$ mode vanishes when defocusing. This must not be surprising since the surficial region, responsible for the $LO$ scattering (see above), is less detected. Last, we have checked that the whole Be-Se signal, including X, remains stable when the laser output is reduced by two orders of magnitude (not shown). Therefore X is intrinsic, and not due to any laser-induced heating at the sample surface. Altogether our experimental results support our assignment of X in terms of a surface PP.

We pursue our discussion on a more quantitative basis by developing a full contour modeling of the Raman signals from the bulk and surface PP's of our three-mode [1×(Zn-Se),2×(Be-Se)] $Zn_{0.67}Be_{0.33}Se$ alloy. We proceed within the general formalism of the linear dielectric response, and calculate the Raman cross sections via the fluctuation-dissipation theorem along the approach of Hon and Faust.[16] The following generic form is eventually derived for the optic vibration of a multi-oscillator ($p$) zincblende alloy,

$$RCS(\omega, x) \propto Im\left\{-\frac{1}{\Delta(\omega,q,x)} \times \left[1 + \sum_p \times C_p(x) \times L_p(\omega,x)\right]^2 + \sum_p \frac{C_p^2(x) \times \omega_p^2(x) \times L_p(\omega,x)}{S_p(x) \times \varepsilon_{\infty,p} \times \omega_p^2}\right\}. \quad (1)$$

The $L_p(\omega, x) = \omega_p^2(x) \times \left(\omega_p^2(x) - \omega^2 - j \times \gamma_p(x) \times \omega\right)^{-1}$ terms describe damped Lorentzian resonances at frequencies $\omega_p(x)$ of the purely-mechanical ($TO_{Zn-Se}, TO_{Be-Se}^{Be}, TO_{Be-Se}^{Zn}$) $p$-oscillators ($p$=1,2,3, correspondingly) of $Zn_{1-x}Be_xSe$. $S_p(x)$ and $C_p(x)$ are the related oscillator strengths and Faust-Henry coefficients, respectively, that scale linearly with the fraction of oscillator $p$, i.e. as $(1-x, x^2, x \cdot (1-x))$ in case of a random Zn↔Be substitution. $\gamma_p(x)$, the phonon damping, is sample-dependent. $\omega_p$ and $\varepsilon_{inf,p}$ are the purely-mechanical $TO$ frequency and the high-frequency relative dielectric constant of the $p$-related parent compound. The ZnSe and BeSe ($S_p, C_p, \omega_p, \varepsilon_{inf,p}$) values are (2.92,−0.7,254.5 cm$^{-1}$,5.75) and (1.77,−0.7,468.5 cm$^{-1}$,5.32), respectively.[15] For $Zn_{0.67}Be_{0.33}Se$, the $\omega_p(x)$ and $\gamma_p(x)$ values for $p$=(1,2,3), given in cm$^{-1}$, are (216.2, 423.5, 468.5) and (10, 40, 40), respectively. Such parameters, which monitor the positions and linewidths at half maximum of the individual Raman lines, are adjusted so as to achieve best contour modelling of the reference $Zn_{0.67}Be_{0.33}Se$ $TO$ backward Raman spectrum. We found that the individual Raman intensities scale as the nominal fractions of oscillators given above (taking $x$=0.33), the sign of a random Zn↔Be substitution in our sample. We emphasize that all above parameters remain fixed in future calculations of the near-forward Raman spectra.

Eq. (1) applies as well to the bulk PP, to the underlying purely-mechanical $TO$ phonon ($q_{inf.}$-regime, $\theta$=180°), to the surface PP or to the $LO$ phonon. Examples are given in **Fig. 2a**. It is just a matter to identify the relevant resonance term $\Delta(\omega, q, x)$, which contains the information on the ($\omega$ vs. $q$) dispersion for each type of mode.

For a bulk PP, the $q$-dependence of the transverse electric field dictated by Maxwell's equations is $\varepsilon_r = \frac{q^2 \times c^2}{\omega^2}$,[1] leading to $\Delta(\omega, q, x) = \varepsilon_r(\omega, x) - \frac{q^2 \times c^2}{\omega^2}$. The classical form $\varepsilon_r(\omega, x) = \varepsilon_{inf.}(x) + \sum_p \varepsilon_{\infty, p} \times S_p(x) \times L_p(\omega, x_p)$ is used, with a linear $x$-dependence for $\varepsilon_{inf.}(x)$. There are two important asymptotic limits. At large $q$, one falls into the $q_{inf.}$-regime, where the bulk PP vanishes into the purely-mechanical $TO$ phonon. Right at $q$=0, the bulk PP becomes $LO$-like.[20]

The surface PP has two main characteristics.[7] First, its wavevector $\vec{q}$ belongs to the air-crystal surface. Second, its electric and magnetic fields vanish on departing from the surface. When using most general forms of such fields, and running Maxwell's equations, one obtains two independent subsets of equations, referring one to a transverse electric (TE) wave and the other to a transverse magnetic (TM) one. In both cases the transverse vector belongs to the surface. As ZnBeSe is non-magnetic, there is no singularity in the permeability by crossing the surface, so that the TE mode cannot propagate. Only the TM mode is allowed. Its dispersion is obtained in two steps. By solving the propagation separately in air and in the crystal, one accesses the related positive terms $\alpha_0 = \sqrt{q^2 - \frac{\omega^2}{c^2}}$ and $\alpha_1 = \sqrt{q^2 - \varepsilon_r \times \frac{\omega^2}{c^2}}$, characteristic of the exponential decays of the fields away from the surface. Besides, the continuity of the electric field at the surface gives $\varepsilon_r = -\frac{\alpha_1}{\alpha_0}$, leading to $\Delta(\omega, q, x) = \varepsilon_r(\omega, x) - \frac{q^2 \times c^2}{\omega^2 - q^2 \times c^2}$.[21] Therefore the surface PP exists only in those frequency domains where $\varepsilon_r$ is negative, i.e. inside $TO - LO$ bands, in contrast with the bulk PP that exists only outside such domains (see above).

The experimentally observed $\omega$ vs. $q$ [1×(Zn-Se),2×(Be-Se)] dispersion, plotted by substituting for $q$ the dimensionless parameter $y = \frac{q \times c}{\omega_1}$, in which $\omega_1$ arbitrarily represents the purely-mechanical $TO$ frequency of ZnSe, are shown in **Fig. 3**. Each $\theta$ value corresponds to an oblique (dashed) line obeying the following relation

$$y = \omega_1^{-1} \times \{n^2(\omega_i, x) \times \omega_i^2 + n^2(\omega_s, x) \times \omega_s^2 - 2 \times n(\omega_i, x) \times n(\omega_s, x) \times \omega_i \times \omega_s \times cos\theta\}^{\frac{1}{2}}, \quad (2)$$

as directly inferred from the wavector conservation rule, being clear that in our Stokes experiment, $\omega_s = \omega_i - \omega$, with the usual meaning for the subscripts. For $n(\omega_i, x)$ and $n(\omega_s, x)$, the refractive indices of the incident and scattered beams in $Zn_{1-x}Be_xSe$, we use the values of Peiris *et al.*[22] The theoretical dispersion curves of the bulk (thick lines) and surface (thin lines) PP's, obtained by solving numerically the relevant $\Delta(\omega, q, x) = 0$ equations, are added for comparison. The reference photon-like (dotted lines) and $(TO, LO)$ phonon-like (explicit labelling) asymptotes are indicated, for clarity. Globally, the Raman frequencies superimpose nicely to the theoretical curves, except for the Zn-Se mode, that remains pinned at a fixed frequency by a parasitical Fano interference (see above).

In view to achieve full contour modelling of the experimental Raman spectra we need to calculate the Raman cross sections of the bulk and surface PP's in their $\theta$-dependence. For doing so we substitute $\theta$ for $q$ (or $y$) in the relevant (bulk or surface) $\Delta(\omega, q, x)$ terms of **Eq. (1)**, using the $\theta$ vs. $q$ (or y) correspondence given by **Eq. (2)**. The theoretical Raman lineshapes of the bulk (thin lines) and surface ($\theta=0°$, dashed line) PP's are superimposed to the corresponding experimental signals in **Fig. 2a**. The agreement is globally excellent, especially when realizing that only one parameter, i.e. $\theta$, monitors all variations in frequency and intensity of the bulk and surface Be-Se doublets (and also of the related Zn-Se modes). This secures, on a quantitative basis, our overall assignment of the BeSe-like PP's.

Generally, the surface PP at the air-crystal surface of a semi-infinite AB crystal is hardly detectable by Raman scattering. This is because it emerges so close to the $LO$ line that it cannot be resolved as a distinct feature.[6] In $Zn_{0.67}Be_{0.33}Se$, the experimental $LO^+_{Be-Se}$ line is shifted to higher frequency with respect to the theoretical prediction (see **Fig. 2a**), due to fine structuring of the Be-Se $LO$ mode induced by the alloy disorder.[15] With this, the $TO - LO$ band is artificially enlarged on its $LO$ side, making the surface PP mode X appear as a distinct Raman feature. In fact, X becomes visible only when the upper Be-Se bulk PP, which emerges nearby, has almost vanished, i.e. at $\theta\sim0°$. We recall that the wavevector of the surface PP is defined within the surface plane, which contradicts the wavector conservation rule at $\theta\sim0°$. Our view is that the activation of X is due to a partial breaking of the wavevector conservation rule induced by the alloy disorder. Such breaking is independently evidenced by the $DA - TA$ and $DA - LA$ bands.

Summarizing, near-forward Raman scattering is used to study the bulk and surface PP's of the three-mode [1×(Zn-Se),2×(Be-Se)] $Zn_{0.67}Be_{0.33}Se$ zincblende alloy, whose short (Be-Se) bond exhibits a distinct percolation-type $TO$ doublet in its conventional backscattering Raman spectra, apparently a universal one among zincblende alloys. This allows to refine the understanding of the PP's in a disordered $A_{1-x}B_xC$ zincblende alloy beyond the current two-mode [1×(A-C),1×(B-C)] MREI approximation.[11] The discussion is supported by contour modeling of the of the bulk and surface PP Raman lineshapes in their $\theta$-dependence within the formalism of the linear dielectric response.[16] Altogether, the backscattering Raman study of $Zn_{0.67}Be_{0.33}Se$ over the past decade,[15] plus its recent study by inelastic neutron scattering,[23] added to the present one by near-forward Raman scattering, complete a state-of-the-art overview of the transverse optic vibrations of an alloy, in their full diversity, throughout the entire Brillouin zone.

Acknowledgement: We would like to thank P. Franchetti for technical assistance in the Raman measurements, C. Jobart for sample preparation, and A. Polian and A.V. Postnikov for useful discussion and careful reading of the manuscript. This work has been supported by the "Fonds Européens de DEvelopment Régional" of Region Lorraine (FEDER project N°. 34619).

**Figure captions**

**Fig. 1:** Unpolarized backward (a, thin curve) and forward (b, thick curve) Raman spectra of the $Zn_{0.67}Be_{0.33}Se$ alloy taken along its [111]-growth axis, as schematically shown. Basic $TO$ differences are indicated by arrows at the bottom. The star marks a Fano interference.

**Fig. 2:** $\theta$-dependent (a) and depth-dependent (b) $LO$-deprived Be-Se Raman spectra taken with the $Zn_{0.67}Be_{0.33}Se$ single crystal by using crossed polarizations of the incident ($\vec{e}_i$) and scattered ($\vec{e}_s$) beams along its $[1\bar{1}0]$ and $[11\bar{2}]$ crystal axis, respectively, as schematically shown. In panel (a) the theoretical bulk (thin lines) and surface ($\theta=0°$, dashed line) PP Be-Se Raman lineshapes together with the $\theta$-insensitive $LO$ one (dotted curve) are superimposed to the experimental curves, for comparison. In panel (b) $\theta$ remains stable (~0°), as testified by the quasi invariance of the lower Be-Se frequency (see the double arrow). The backscattering ($\theta=180°$) $TO$ frequencies are emphasized in both panels (dashed lines), for reference purpose.

**Fig. 3:** Theoretical bulk (thick lines) and surface (thin lines) $\omega$ vs. $y$ ($=\frac{qc}{\omega_1}$) PP dispersion curves obtained for the three-mode [1×(Zn-Se),2×(Be-Se)] $Zn_{0.67}Be_{0.33}Se$ alloy by solving numerically the relevant $\Delta(\omega,q,x) = 0$ equations. The experimentally accessible $\omega$ vs. $y$ relations by near-forward Raman scattering at selected $\theta$ values (quasi parallel oblique dashed lines) are superimposed. The Raman frequencies are quoted (circles), spoilt by a fixed error corresponding to the Raman linewidth at half maximum. Useful ($TO$,$LO$) phonon-like (explicit labeling) and [surface: (a), bulk: (b)-(c)] photon-like (dotted lines) asymptotes are indicated. The sign of $\varepsilon_r$ is specified on the right.

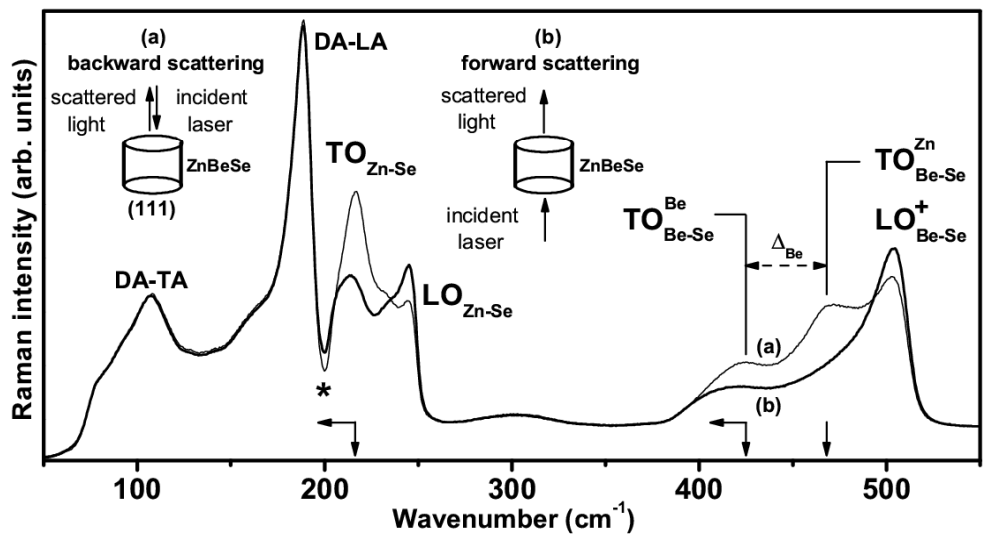

**Figure 1**

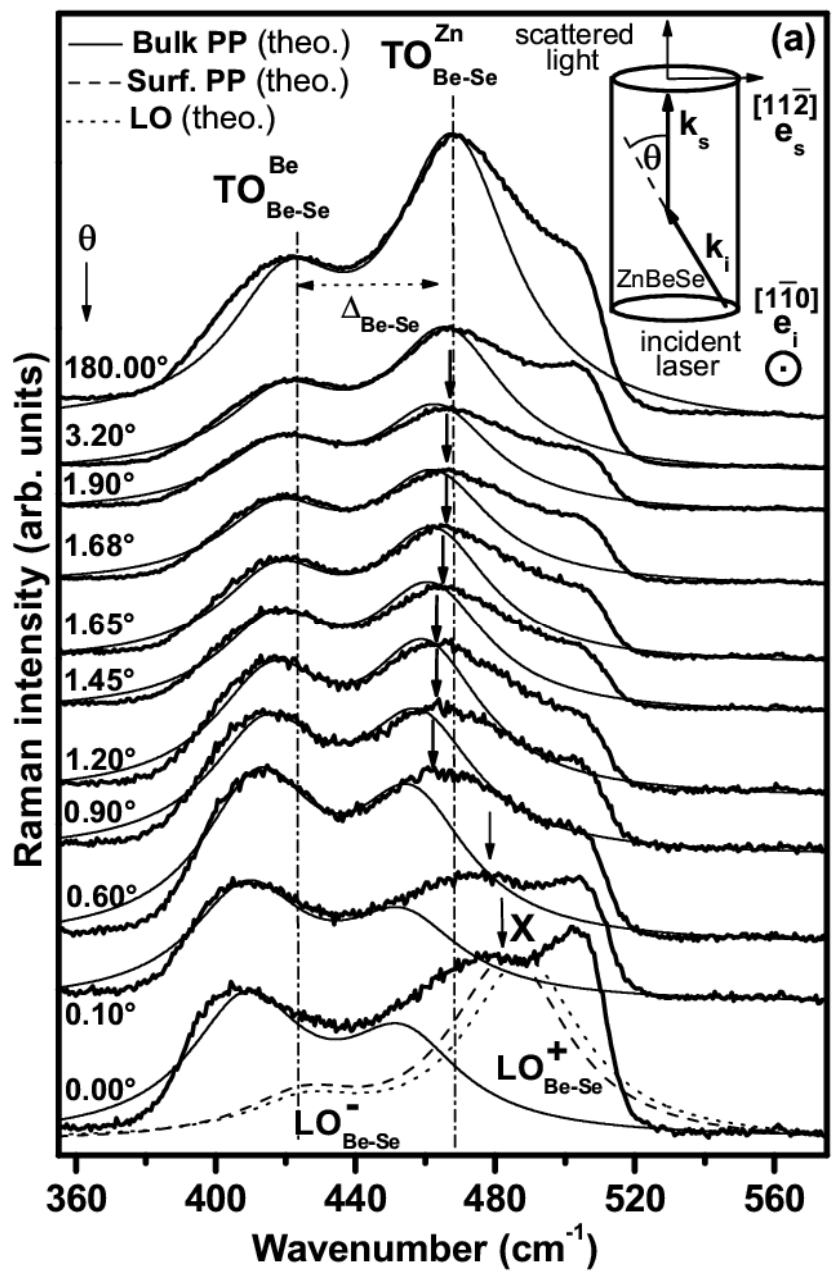

**Figure 2a**

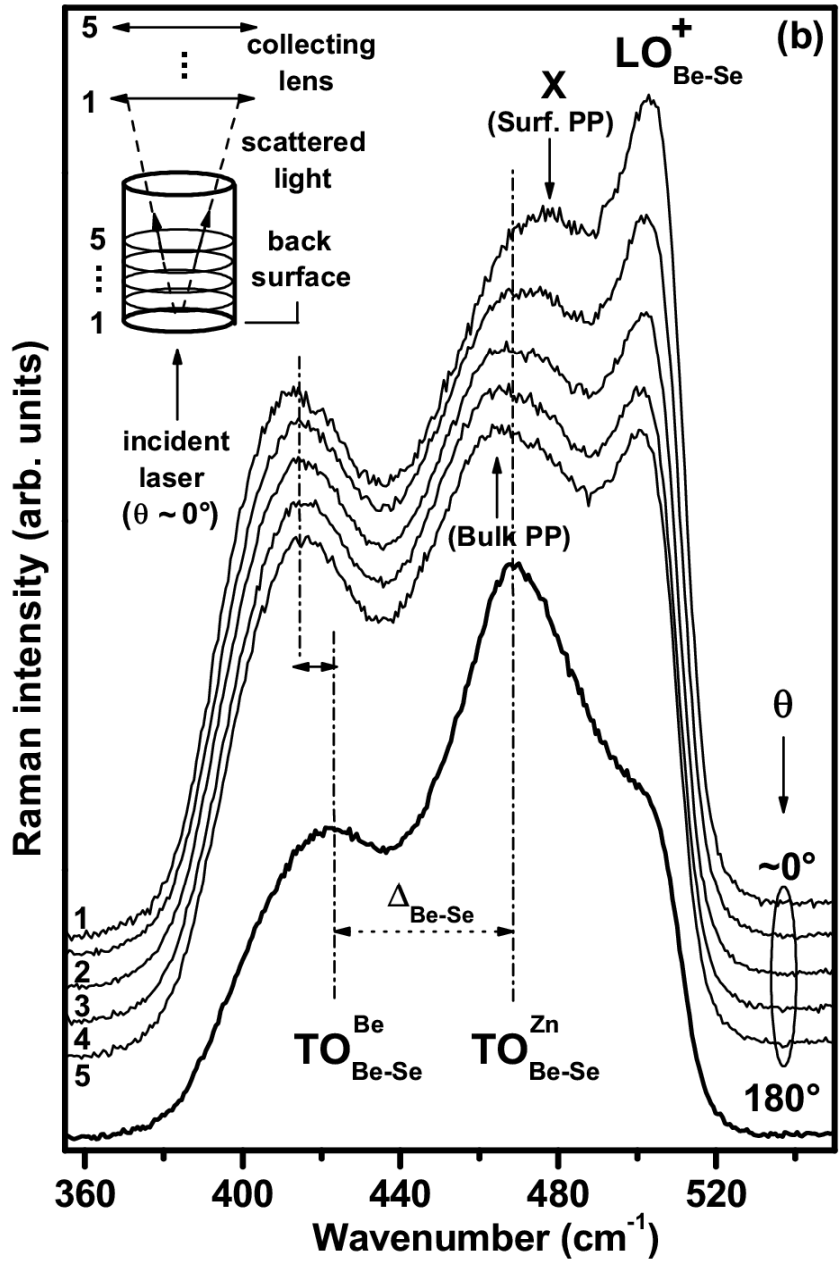

**Figure 2b**

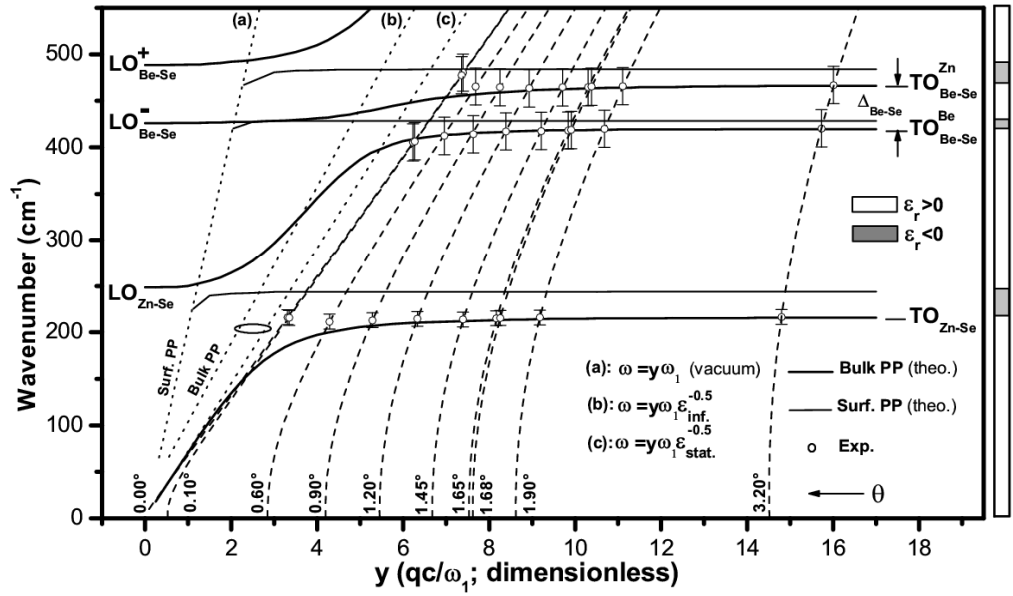

**Figure 3**